\documentclass[useAMS,usenatbib]{mn2e}
\usepackage{graphicx}
\usepackage{natbib}

\title[Dissecting the Galactic Anticenter]{Chemo-dynamical properties of the Anticenter Stream: a surviving disc fossil from a past satellite interaction.}
\author[Laporte et al.]{
\parbox[t]{\textwidth}{Chervin F. P. Laporte$^{1}$\thanks{E-mail:cfpl@uvic.ca}\thanks{CITA National Fellow}, Vasily Belokurov$^{2}$, Sergey E. Koposov$^{3,2}$, Martin C. Smith$^{4}$,Vanessa Hill$^{5}$
}
\\
\\
$^{1}$Department of Physics \& Astronomy, University of Victoria, 3800 Finnerty Road, Victoria BC, Canada V8P 5C2\\
$^{2}$Institute of Astronomy, University of Cambridge, Madingley road, CB3 0HA, UK\\
$^{3}$McWilliams Center for Cosmology, Carnegie Mellon University, 5000 Forbes Ave, 15213, USA\\
$^{4}$Key Laboratory for Research in Galaxies and Cosmology, Shanghai Astronomical Observatory, Chinese Academy of Sciences, 
80 Nandan Road,\\ Shanghai 200030, China\\
$^{5}$Laboratoire Lagrange, Universit\'e C\^ote d’Azur, Observatoire de la C\^ote d’Azur, Boulevard de l’Observatoire, CS 34229, 06304, Nice, France\\
}
\date{Accepted . Received ; in original form }
\pubyear{2018}
\begin{document}
\label{firstpage}
\pagerange{\pageref{firstpage}--\pageref{lastpage}}
\maketitle
% Abstract of the paper
\begin{abstract}
%It has been suggested that the Anticenter Stream (ACS) may be the remains of a stellar disc tidal tail excited by the interaction with the Sagittarius dwarf galaxy. 

Using Gaia DR2, we trace the Anticenter Stream (ACS) in various
stellar populations across the sky and find that it is kinematically
and spatially decoupled from the Monoceros Ring. Using stars from {\sc lamost} and {\sc segue}, we show that the ACS is systematically more metal-poor than Monoceros by $0.1$ dex with indications of a narrower metallicity spread. Furthermore, the ACS is predominantly populated of old stars ($\sim 10\,\rm{Gyr}$), whereas Monoceros has a pronounced tail of younger stars ($6-10\, \rm{Gyr}$) as revealed by their cumulative age
distributions. Put togehter, all of this evidence support predictions from simulations
of the interaction of the Sagittarius dwarf with the Milky Way, which
argue that the Anticenter Stream (ACS) is the remains of a tidal tail
of the Galaxy excited during Sgr's first pericentric passage after
it crossed the virial radius, whereas Monoceros consists of the
composite stellar populations excited during the more
extended phases of the interaction. We suggest that the ACS can be
used to constrain the Galactic potential, particularly its flattening,
setting strong limits on the existence of a dark disc. Importantly,
the ACS can be viewed as a stand-alone fossil of the chemical
enrichment history of the Galactic disc.
\end{abstract}
% Select between one and six entries from the list of approved keywords.
% Don't make up new ones.
%%%%%%%%%%%%%%%%%%%%%%%%%%%%%%%%%%%%%%%%%%%%%%%%%%
%%%%%%%%%%%%%%%%% BODY OF PAPER %%%%%%%%%%%%%%%%%%
\section{Introduction}
Signs of vertical disc perturbations to the disc have been known from the
distribution of the neutral hydrogen gas since the 1950s
\citep{burke57}. In recent years, interest in disc quakes -
particularly in the stellar component - has been reinvigorated with
the discovery of spatial and kinematic North/South asymmetries around
the solar neighbourhood
\citep{widrow12,williams13,carlin13,carrillo18,schoenrich18}. As
revealed by star count studies, these asymmetries take their most
dramatic form in the outer edge of the Galaxy where the self-gravity
of the disc is at its weakest. Amongst the various tributaries of the
Galactic Anticenter, we count the Monoceros Ring \citep{newberg02,slater14} and its substructured content, namely the Anticenter Stream (ACS) and
Eastern Banded Structure \citep{grillmair06} that most visibly stand
out in main-sequence (MS) and main-sequence turn-off (MSTO) star
counts. Although initially thought to be part of the remains of a
torn-apart accreted dwarf galaxy \citep{penarrubia05}, recent
theoretical \citep{gomez15b,laporte18b,laporte19a} and observational
studies favour an excited disc origin for these structures as
supported by the kinematics \citep{deboer17, deason18} and stellar
populations \citep[e.g. see][]{price-whelan15, sheffield18} with
manifestly disc-like properties. 

Recently, \cite{laporte19a} suggested a re-interpretation of the Anticenter Stream as the remnant of a tidal tail (``feather'') excited by the Sgr dwarf galaxy shortly after virial radius crossing. A falsifiable predictions of this scenario is that the stellar populations of Monoceros and the ACS should show differences in metallicity and age distributions. In particular, because the ACS got excited through resonant processes with the reaction of the Milky Way's dark matter halo with Sgr, it must have decoupled itself from the rest of the disc and should hold predominantly old stars and very few young ones. With the current synergy between legacy spectroscopic surveys ({\sc segue}, {\sc lamost}, {\sc apogee}) and the second data release from Gaia, it is now possible to dissect the Anticenter to much greater detail and test the tidal tail remnant hypothesis with chemistry and dynamics. This is the aim of the following contribution. In Section 2, we discuss our target selection and masks in the Monoceros Complex and cross-matches to the aforementioned spectroscopic surveys. In Section 3, we dissect the Monoceros Ring and ACS in metallicity, age and abundance. We discuss our main findings and conclude in Sections 4 and 5.
\begin{figure}
\includegraphics[width=0.5\textwidth,trim=13mm 13mm 20mm 10mm, clip]{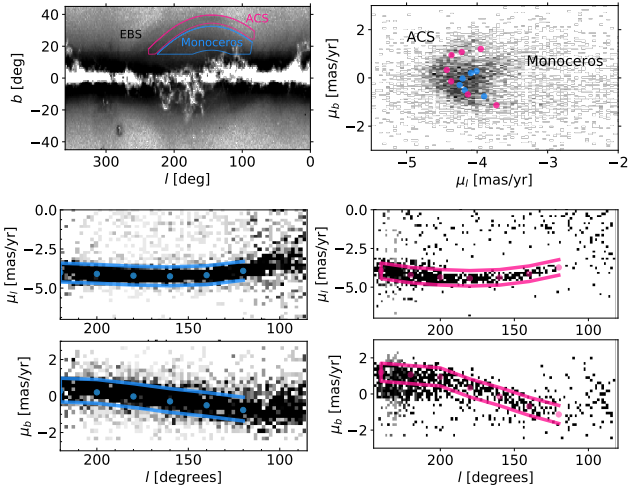}\\
\caption{{\it Top left:} MS/MSTO map of the Anticenter in Galactic coordinates $(l,b)$ with selected spatial footprints for the ACS and Monoceros Ring. ACS sits above the Monoceros Ring as a long collimated thin structure. {\it Top right:} Latitudinal against longitudinal proper motions for RC stars. The median proper motion tracks for the ACS and Monoceros are shown by the magenta and blue dots. {\it Middle left:} Normalised histogram for longitudinal proper motions $\mu_l$ as a function of $l$ for ``ACS field" stars. Proper motion masks and median $\mu_l$ within them are shown in magenta. {\it Middle right:} Normalised histogram for longitudinal proper motions $\mu_{l}$ as a function $l$ for ``Monoceros field" stars. Proper motion masks and median $\mu_l$ within them are shown in blue. {\it Bottom left:} Normalised histogram for latitudinal proper motions $\mu_b$ as a function of $l$ for ``ACS field" stars. Note that this mask avoids the contamination from Monoceros situated below $\mu_{b}\sim1 \rm{mas/yr}$ for $l>200$ as the ACS and Monoceros overlaps spatially in those zones due to its reconnection with the midplane, yet show remarkably different kinematics. Proper motion masks and median $\mu_b$ within them are shown in magenta. {\it Bottom right:} Normalised histogram for latitudinal proper motions $\mu_b$ as a function of $l$ for ``Monoceros field" stars. Proper motion masks and median $\mu_b$ within them are shown in blue. }
\label{fig:selection}
\end{figure}

\begin{figure}
\includegraphics[width=0.5\textwidth,trim=45mm 30mm 52mm 20mm, clip]{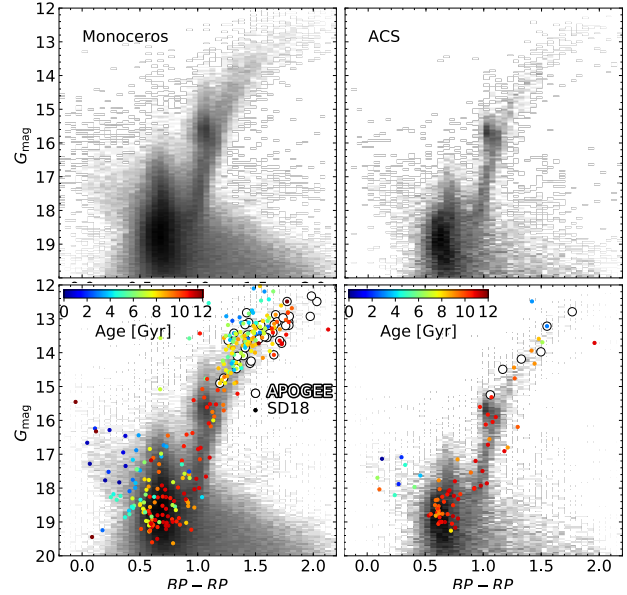}\\
\caption{ {\it Top panels:} CMDs for Monoceros and ACS fields (left and
  right panels respectively). The kinematic and spatial selection
  picks up a well defined RGB, RC all the way down to a MSTO/MS. The
  excess of blue stars may be a population of blue stragglers
  (BS). {\it Bottom panels:} Stars identified in the \citet{sanders18} catalog are shown as points colour-coded as a function of age. {\sc apogee} cross-matched stars are shown as open white-filled black circles.}
\label{fig:CMD}
\end{figure}

\section{Target selection}
For our study, we make use of the GDR2 proper motions and parallaxes
to identify the likely ACS and Monoceros members. We correct magnitudes and colours for extinction by using the \cite{schlegel98} dustmap and assuming $A_{X}=A_{0} k_{X}$, where $X$ designates the passband and $k_{X}$ the first extinction coefficient of the relation used by \cite{danielski18} adapated to the Gaia passabands assuming that $A_{0}=3.1E(B-V)$ \citep{babusiaux18}. To guide our initial spatial search, we begin by selecting the main-sequence/main-sequence
turn-off (MS/MSTO) by requiring $0.5<BP-RP<0.8$ and $18<G_{0}<20$ and
$\varpi<0.1$. We then convert the coordinates to a new system
approximately aligned with the Anticenter by rotating the celestial
equator to the great circle with a pole at (l,b) = (325.00,
67.4722). The ACS and Monoceros stars are drawn from the spatial masks
shown in Galactic coordinates in the top left panel of
Figure~\ref{fig:selection}. Moreover, in the top right panel of the
Figure, using the Red Clump (RC) stars (selected with the cuts
$15.5<G_{0}<15.9$, $1.0<BP-RP<1.1$ and $\varpi<0.1$), we demonstrate
that the Monoceros Ring and the Anticenter stream not only have
distinct spatial distributions but also differ kinematically. This is
evidenced by the bifurcating pattern in $(\mu_l,\mu_b)$ space for
which we also present median proper motion tracks in magenta (blue)
for the ACS (Monoceros) respectively. This two-horned structure
confirms some of the earlier suggestions of \cite{deboer17} based on
the Gaia DR1-SDSS astrometric analysis.

We proceed by using Gaia DR2 to identify high-fidelity candidate stars
belonging to the ACS and the Monoceros regions.  We make use of the
parallax cut $\varpi<0.1$ as well as the proper motion masks shown in
Figure~\ref{fig:selection}. The two bottom rows of the Figure give
column-nornalized RC density in the space of $\mu_b$ and $\mu_l$
proper motions as a function of Galactic longitude $l$ for Monoceros
(ACS) in the left (right). The proper motion masks - highlighted by
the magenta and blue boxes respectively - are chosen to include the
highest-density signal at each $l$. Figure~\ref{fig:CMD} presents CMDs
for both fields and only displays stars that have passed the proper
motion and parallax cuts described above. Readily identifiable in the
Figure are several familiar stellar populations: MS/MSTO, RC and red
giant branch (RGB), thus confirming that our selection picks up
bona-fide stars associated with the two individual well-defined
structures. Moreover, note that the Hess diagram of the Monoceros Ring
is much broader compared to the ACS, which signals a larger mix of
metallicities and ages and possibly line-of-sight distances.

%These masks will ensure us that we make no strong assumptions on
%distance, metallicity or ages of the stellar populations, as opposed
%to using a colour-magnitude (CMD) mask with isochrones which may bias
%results.

\section{Chemical and age decomposition of Monoceros and the ACS}

\begin{figure}
\includegraphics[width=0.5\textwidth,trim=50mm 38mm 70mm 62mm, clip]{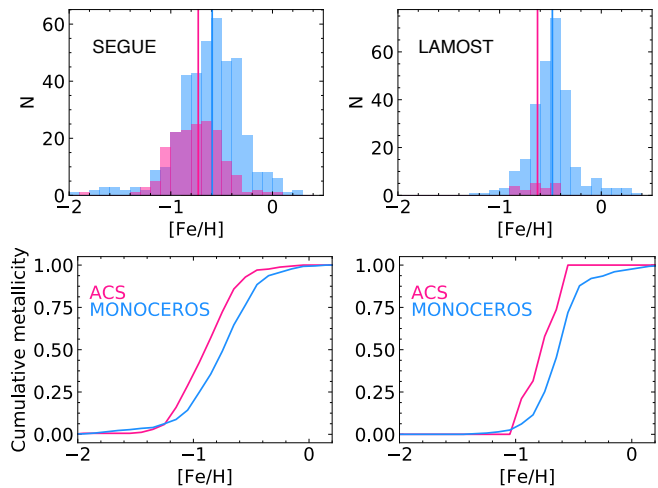}\\
\caption{ {\it Left:} Metallicity distribution for stars cross-matched
  with {\sc segue} for the ACS and Monoceros Ring in magenta and blue
  respectively. The median and spreads in metallicity $([\rm{Fe/H}],
  \sigma[\rm{Fe/H}])$ are $(-0.73, 0.26)$ and $(-0.59,0.32)$ for the
  ACS and Monoceros Ring respectively. {\it Right:} Metallicity
  distribution for stars cross-matched with {sc lamost} for the ACS
  and Monoceros Ring in magenta and blue respectively. The median and
  spreads in metallicity $([\rm{Fe/H}], \sigma [\rm{Fe/H}])$ are
  $(-0.62, 0.14)$ and $(-0.48,0.22)$ for the ACS and Monoceros Ring
  respectively.}
\label{fig:mdf}
\end{figure}

\subsection{Metallicity distributions}
The {\sc lamost} DR4 survey \citep{luo15} provides a good coverage of the Anticenter
region, with a large number of spectra measured across both structures
without a strong metallicity bias \citep[see e.g.][]{Yanny09}. {\sc
  segue} fares similarly well and has also the advantage of reaching
down to the main-sequence and the turn-off at larger distances. This
is particularly advantageous to analyse differences in age
distributions between Monoceros and the ACS for which the MSTO is
sensitive to. Note however that {\sc segue} includes a sub-dominant
target category (F sub-dwarfs) biased against metal-richer
stars. Furthermore, in order to avoid any source of confusion we only
analyse stars in the range $115<l<175$ as these regions separate most
clearly between the ACS and Monoceros fields both spatially and
kinematically (see Figure~\ref{fig:selection}). We select spectra with
signal-to-noise ratios of SNR$\geq10$ for {\sc lamost} a SNR$\geq20$
for {\sc segue}. Figure~\ref{fig:mdf} shows metallicity distributions
for the likely Monoceros and the ACS members from our cross-matches
with {\sc lamost} and {\sc segue}. Although the numbers differ from
one survey to another, both spectroscopic samples reveal similar
systematic trends. The median metallicity of the ACS is consistently
lower than that of Monoceros by some $\Delta[Fe/H]\sim 0.1 \,
\rm{dex}$.

% Moreover, % Monoceros shows tentative signs of a larger metallicity
% spread than the ACS by $\sim0.1\, \rm{dex}$.
% applied SNR>20 cut for SEGUE. changing to SNR>10 didn't change anything either
% in order to get enough stars I had to push to SNRg>10 in LAMOST

\subsection{Age distributions}

\begin{figure}
\includegraphics[width=0.5\textwidth,trim=0mm 0mm 0mm 0mm, clip]{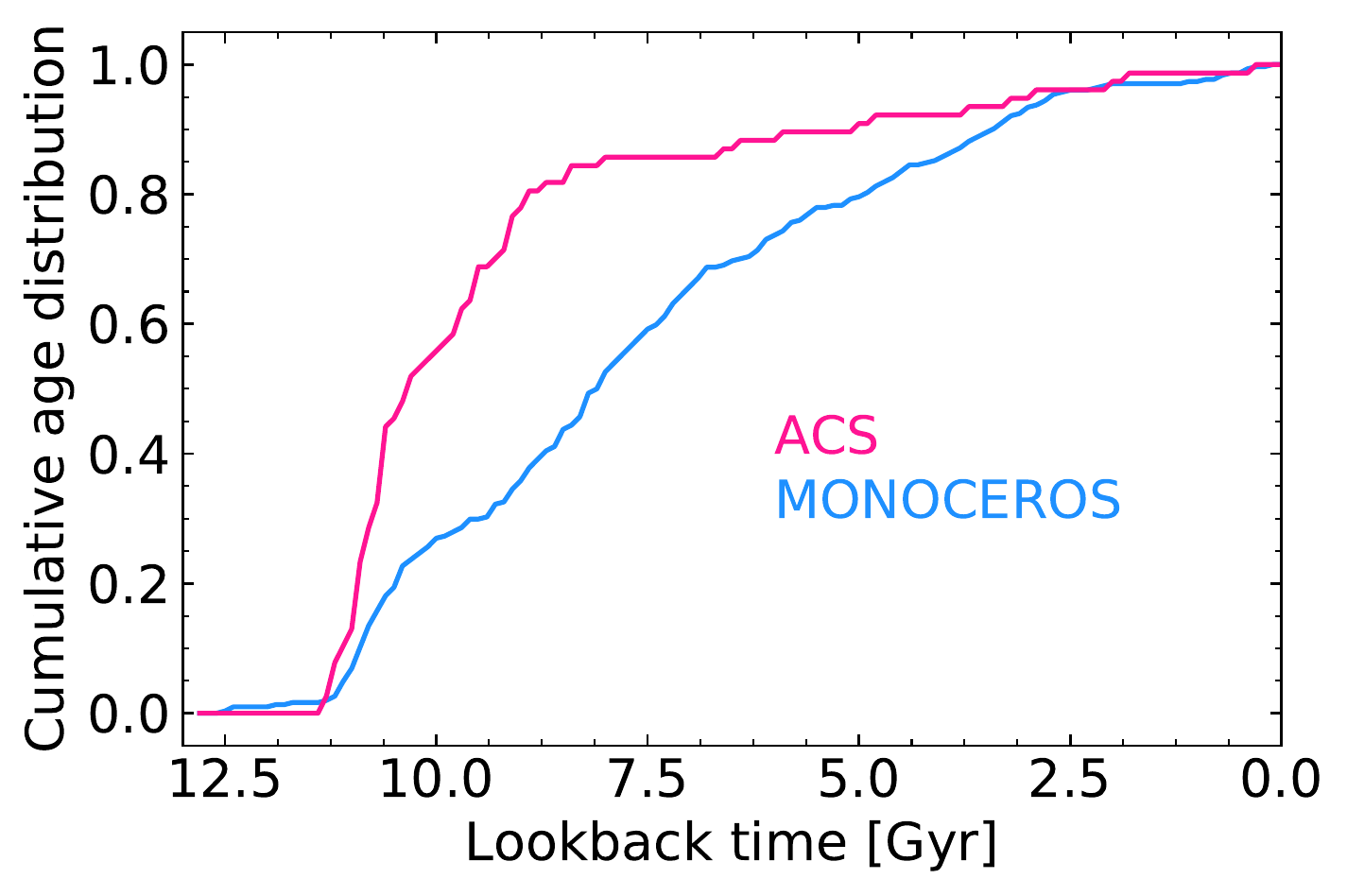}\\
\caption{Cumulative age distribution of stars cross-matched with the
  \citet{sanders18} catalog. The ACS shows a rapid increase in its
  cumulative distribution with about 80\% of the stars being older
  than 10 Gyr, whereas the Monoceros Ring shows a much more steady
  formation of stars.}
\label{fig:ages}
\end{figure}

\begin{figure}
\includegraphics[width=0.5\textwidth,trim=0mm 0mm 0mm 0mm,
  clip]{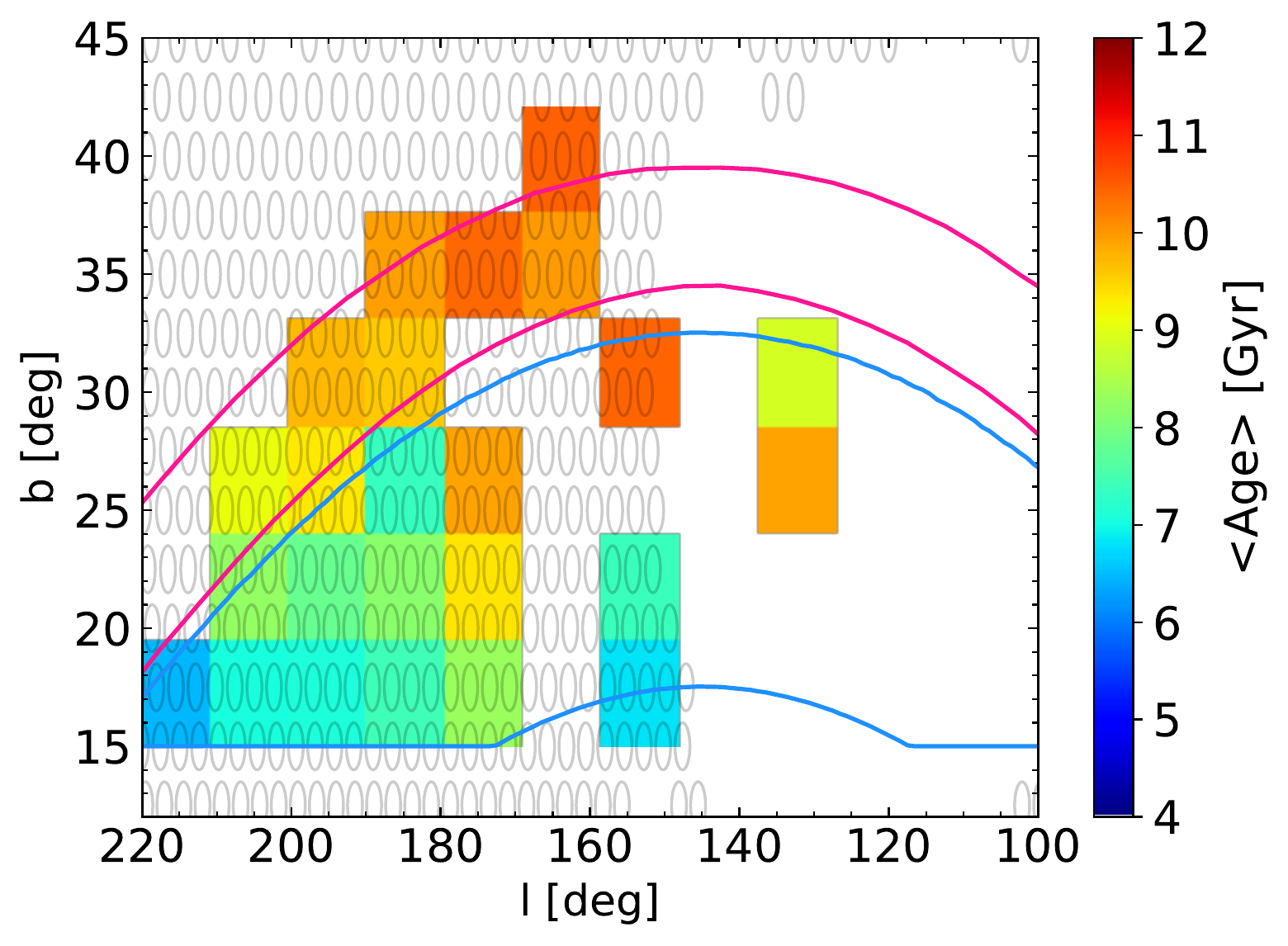}\\
\caption{Spatial map of star median ages. Footprints of Monoceros and the ACS are shown in blue and magenta respectively. {\sc weave} high-resolution footprint (Jin et al. in prep) is overlaid in black open circles.}
\label{fig:agemap}
\end{figure}

To study the star-formation histories of the two structures, we
cross-match the candidate stars identified above with the catalog of
stellar ages computed by \cite{sanders18}. Figure~\ref{fig:ages} shows
the cumulative age distributions for the ACS and Monoceros
regions \footnote{We have also checked that our results remain
  unchanged when focusing only on the MSTO stars, which would give the
  best age estimates compared to the RC and RGB}. The difference
between the two structures are remarkable, with the ACS being
predominantly composed of older stars $\tau_{{\rm ACS}}>10\,\rm{Gyr}$
whereas Monoceros possessing a more steady, gradual star-formation
history. We checked that there was no correlation between age and metallicity in our subset of \cite{sanders18} cross-matched stars and that the distances are consistent with the structures ($d\sim 10\,\rm{kpc}$). In the encounter scenario \citep[see][]{laporte19a}, the ACS is a group of stars located in a tidal tail of the Galactic disc which
gets decoupled from the rest of the disc and propelled to larger
heights from midplane after first pericentric passage of a massive
satellite (e.g. Sgr), whereas Monoceros consists of stellar
populations in the flared and corrugated outer disc which was
gradually built up through a succession of encounters, allowing it to
replenish itself with younger stars as the star formation proceeded.
Figure~\ref{fig:agemap} presents a spatial median age map of the Anticenter
region. We find that the ACS is systematically older than the
Monoceros Ring which hosts plenty of intermediate age stars (5-9 Gyr).
Note a sharp age boundary between the two structures matching the
location of the density transition
(c.f. Figure~\ref{fig:selection}).% Future surveys should in principlebe able to reveal a full age coverage of the Anticenter with different tracers.

\subsection{ {\sc apogee} chemical abundances for the ACS and Monoceros}

Despite its pencil-beam nature, the {\sc apogee} survey \citep{majewski17} covers parts of both the ACS and the Monoceros Ring. This allows us to acquire
alpha-element abundances for our candidate stars through
cross-matching catalogs. This gives us a few candidate stars which
fall within the RGB/RC as shown in Figure 2. In Figure 6, we show the
locations of our {\sc apogee} cross-matched stars in the the space of
$([Mg/Fe], [Fe/H])$. Not surprisingly, these stars belong to the
low-$\alpha$ sequence with $0<[Mg/Fe]<0.15$ and low metallicity
$-0.8<[Fe/H]<0.3$, commonly known as the chemical ``thin disc'', which
confirms that the ACS and Monoceros Ring are not tidal debris from
accretion events but truly extensions of the outer disc. This is not
surprising as other/similar structures of the Anticenter have also
recently been confirmed to be chemical thin-disc material through
abundance measurements \citep[e.g. see][]{bergemann18} and stellar
populations content \citep{price-whelan15, sheffield18}.

\begin{figure}
\includegraphics[width=0.5\textwidth,trim=0mm 0mm 0mm 0mm,
  clip]{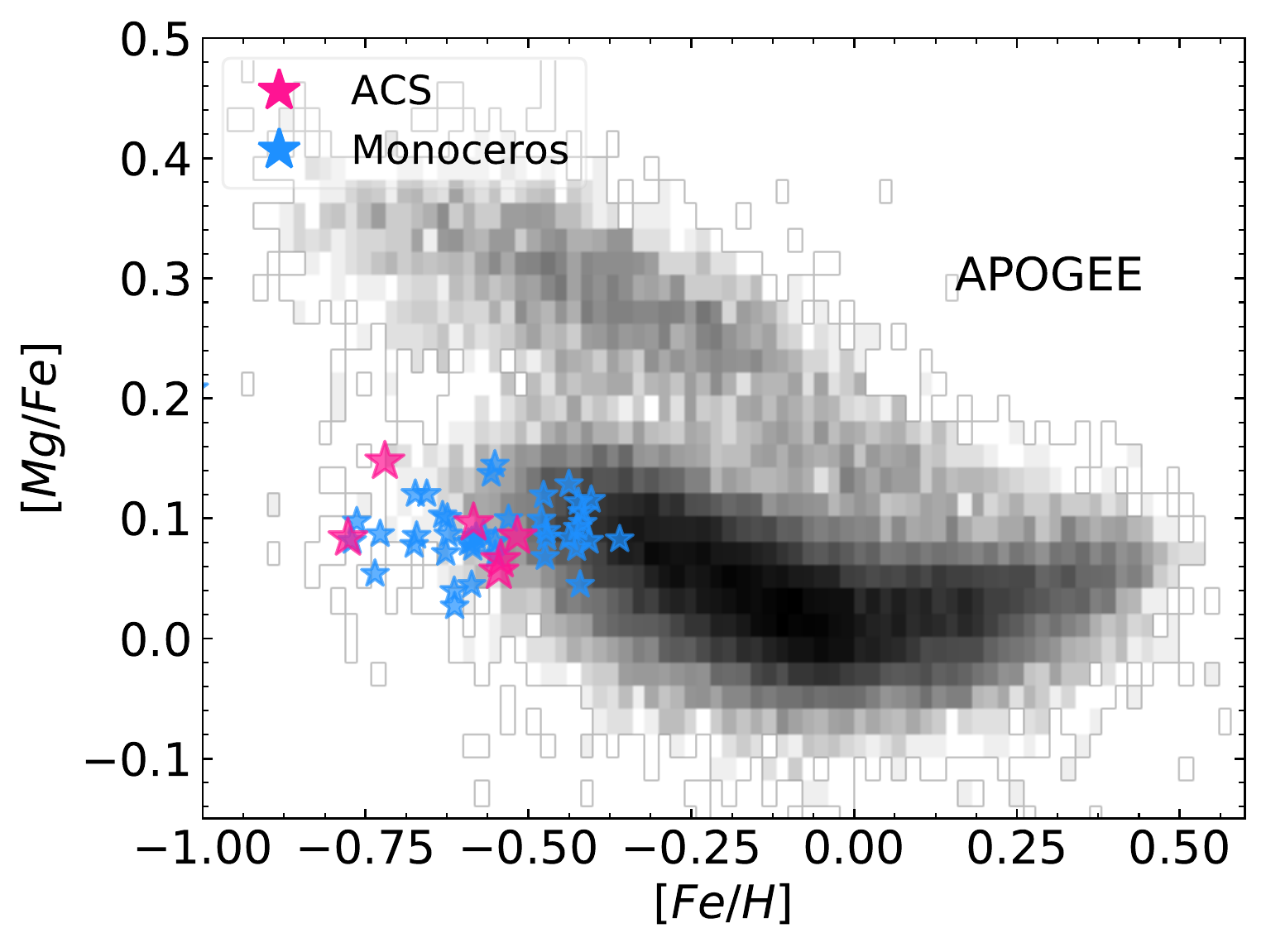}\\
\caption{Magnesium abundance versus metallicity of stars in ACS and
  Monoceros within the proper motion masks defined in Section 1
  (magenta and blue star symbols respectively). Despite the pencil
  beam nature of the {\sc apogee} survey, a few stars associated with
  Monoceros and the ACS are picked up and show a clear low-alpha
  abundance sequence consistent with the chemical ``thin disc". The
  full {\sc apogee} disc $[Mg/Fe]$ vs $[Fe/H]$ map is displayed in
  grey-scale for comparison.}
\label{fig:alpha}
\end{figure}

\section{Discussion}

By using a combination of astrometric, photometric and spectroscopic
information, we were able to dissect the Monoceros Ring and the ACS in
the space of kinematics, metallicities, $\alpha$-abundances and
ages. This allowed us to explore and confirm a falsifiable prediction
for their respective formation mechanisms as presented in
\citep{laporte19a}, namely that the ACS is the remnant tidal tail of
the MW disc which formed through a resonant interaction with the Sgr
dwarf galaxy. The ACS was kicked up shortly after the dwarf's crossing
of the Galaxy's virial radius during one of the first pericentric
passages. The ACS excitation resulted in a strong decoupling from the
Galactic midplane, leading to a sudden shutdown of star formation as
compared to the rest of the disc. This yielded the observed striking
difference in the cumulative age distributions between the Monoceros
and the ACS. 

We note that several structures in the outer disc have also been
identified. These include the EBS \citep{grillmair06} and the more
distant TriAnd clouds \citep{rocha-pinto04}. A similar analysis could
in principle be pursued, in particular for the TriAnd, which lies at a
larger distance. This is particularly interesting as these structures
may represent a fossil record of the formation history of the outer
disc.  Via modelling of such structures one can hope to time the
impact events, putting strong constraints on the orbital mass-loss
history of the Sgr dwarf galaxy. Our analysis argues that it may be
possible to use chemistry and age dating important events in the
lifetime of the Galactic disc. The decoupled nature of the structures
analysised here - the ACS and the Monoceros - is of particular
interest for chemo-dynamical models of the Galaxy
\citep[e.g.][]{chiappini97,schoenrich09}. 

Given the disc nature of the ACS, and the relatively simple dynamics
of ``feathers'' \citep{laporte19a}, this structure may also be used
for constraining the flattening of the Galactic potential at large
radii, thus setting strong limits on alternative dark matter models or
the existence of a dark disc \cite{read08}, however this is beyond the
scope of this contribution and will be presented elsewhere.

\section{Conclusion}
In this work, we took full advantage of the synergy between Gaia DR2,
{\sc segue}, {\sc lamost} and {\sc apogee} to show that:
\begin{enumerate}
    \item The ACS and Monoceros Ring are spatially and kinematically
      separate structures.
    \item The ACS is on average more metal-poor than the Monoceros
      Ring, by $\geq 0.1 \,\rm{dex}$, with hints of a smaller spread
      in metallicity (though this could perhaps be accounted by distance spreads too).
    \item The ACS and Monoceros Ring are both part of the chemically
      thin-disc due to their low magnesium abundances, with
      $0.0<[Mg/Fe]<0.15$.
    \item The ACS has predominantly old stellar populations with 80\%
      of the having an age $>10 \,\rm{Gyr}$. This taken with its
      physical and kinematic decoupling from the rest of the disc,
      supports the hypothesis that this group of stars is a
      ``feather'', i.e. the remnant of a tidal tail excited by a
      satellite encounter such as that with the Sgr dwarf described in
      \citet{laporte19a}. In this model, the ACS is extracted from the
      disc during the dwarf's first passage after virial radius
      crossing, and no longer forms stars.
%    \item The Hess diagrams for the Monoceros Ring seems also broader than that of the ACS, suggesting the fact that it would consist of a wider mix of metallicities and ages.  
    \item The Monoceros Ring shows a steady cumulative age
      distribution suggesting that it belongs to main body of the disc
      which has been gradually flared and corrugated as a result of
      the multiple passages of Sgr and populated by stars of different
      ages as star formation continued.
\end{enumerate}

As an outlook into the future, surveys such as {\sc weave}, {\sc sdss
  v}, {\sc 4most} and {\sc psf} will pave the road to a full coverage
of the Anticenter. These surveys will not only provide radial
velocities for a full characterisation of the the phase-plane spiral
in the outer disc as predicted by numerical models of the interaction
of Sgr with the Milky Way \citep{laporte18b,laporte19c} but will also
allow for a more detailed chemical dissection of the Anticenter. In
particular, the latter will provide a window into the fossil record of
the Galactic disc's formation.

\section*{Acknowledgements}
%CL thanks the David W. Hogg for hosting him during the Gaia Data release meeting in NYC where some of this work was partially started. 
This work has made use of data from the European Space Agency (ESA) mission Gaia (https://www.cosmos.esa.int/gaia), processed by the Gaia Data Processing and Analysis Con- sortium (DPAC, https://www.cosmos.esa.int/web/gaia/ dpac/consortium). Funding for the DPAC has been pro- vided by national institutions, in particular the institutions participating in the Gaia Multilateral Agreement. This work made use of {\tt numpy, scipy} and {\tt matplotlib} \citep{numpy, scipy, hunter07} as well as the {\tt astropy} package \citep{astropy1, astropy2}. This paper made use of the Whole Sky Database (wsdb) created by S. Koposov and maintained at the Institute of Astronomy, Cambridge by S. Koposov, V. Belokurov and W. Evans with financial support from the Science \& Technology Facilities Council (STFC) and the European Research Council (ERC). CL \& VB acknowledge support in part by KITP with support from the Heising-Simons Foundation and the National Science Foundation (grant No. NSF PHY-1748958). SK is partially supported by NSF grant AST-1813881 and Heising-Simons foundation grant 2018-1030. MCS acknowledges financial support from the National Key Basic Research and Development Program of China (No. 2018YFA0404501) and NSFC grant 11673083. CL thanks Kathryn V. Johnston, Jorge Pe\~narrubia, Julio F. Navarro and Isabel M. E. Santos-Santos for useful discussions.
\bibliographystyle{mn2e}
\bibliography{master2.bib}{}
%%%%%%%%%%%%%%%%%%%%%%%%%%%%%%%%%%%%%%%%%%%%%%%%%%
%%%%%%%%%%%%%%%%%%%% REFERENCES %%%%%%%%%%%%%%%%%%
% The best way to enter references is to use BibTeX:
%\bibliographystyle{mnras}
%\bibliography{example} % if your bibtex file is called example.bib
% Alternatively you could enter them by hand, like this:
% This method is tedious and prone to error if you have lots of references
%\begin{thebibliography}{99}
%\end{thebibliography}
%%%%%%%%%%%%%%%%%%%%%%%%%%%%%%%%%%%%%%%%%%%%%%%%%%
%%%%%%%%%%%%%%%%% APPENDICES %%%%%%%%%%%%%%%%%%%%%
%%%%%%%%%%%%%%%%%%%%%%%%%%%%%%%%%%%%%%%%%%%%%%%%%%
% Don't change these lines
\bsp	% typesetting comment
\label{lastpage}
\end{document}